\documentclass[natbib]{emulateapj}

\shorttitle{ Possible states for a
  neutron star in {\lsi}} \shortauthors{A.~Papitto, D. F. Torres, \&
  N. Rea}

\usepackage{natbib}
\usepackage[usenames]{color}

\usepackage{graphicx}

\def\lsi{LS~I~+61$^{\circ}$303}
\def\ltsima{$\; \buildrel < \over \sim \;$}
\def\simlt{\lower.5ex\hbox{\ltsima}}
\def\gtsima{$\; \buildrel > \over \sim \;$}
\def\simgt{\lower.5ex\hbox{\gtsima}}

\makeatletter

\newcommand{\be}{\begin{equation}}
\newcommand{\ee}{\end{equation}}
\newcommand{\ba}{\begin{eqnarray}}
\newcommand{\ea}{\end{eqnarray}}

\newcommand{\Rmnum}[1]{\expandafter\@slowromancap\romannumeral #1@}
\makeatother

\begin{document}

\title{ Possible changes of state and relevant timescales for a
  neutron star in {\lsi}}

\author{A.~Papitto\altaffilmark{1}, D.~F.~Torres\altaffilmark{1,2} and N.~Rea\altaffilmark{1}}
\affil{$^1$ Institut de Ci\`encies de l'Espai (IEEC-CSIC)
Campus UAB, Fac. de Ci\`encies, Torre C5 parell, 2a planta
08193 Barcelona,Spain \\
$^2$ Instituci\'o Catalana de Recerca i Estudis Avan\c{c}ats (ICREA), 08010 
Barcelona, Spain}

\begin{abstract}

The properties of the short, energetic bursts recently observed from
the $\gamma$-ray binary {\lsi}, are typical of those showed by
high magnetic field neutron stars, and thus provide a strong
indication in favor of a neutron star being the compact object in the
system.  Here, we discuss the transitions among the states accessible
to a neutron star in a system like {\lsi}, such as the ejector,
propeller and accretor phases, depending on the NS spin period,
magnetic field and rate of mass captured. We show how the observed
bolometric luminosity ($\simgt {\rm few}\times10^{35}$ erg s$^{-1}$),
and its broad-band spectral distribution, indicate that the compact
object is most probably close to the transition between working as an
ejector all along its orbit, and being powered by the propeller effect
when it is close to the orbit periastron, in a so-called {\it
  flip-flop} state. By assessing the torques acting onto the compact
object in the various states, we follow the spin evolution of the
system, evaluating the time spent by the system in each of them.  Even
taking into account the constraint set by the observed $\gamma$-ray
luminosity, we found that the total age of the system is compatible
with being $\approx$ 5--10 kyr, comparable to the typical spin-down
ages of high-field neutron stars. The results obtained are discussed
in the context of the various evolutionary stages expected for a
neutron star with a high mass companion.

\end{abstract}


\section{Introduction}

{\lsi} is one the few high-mass X-ray binaries (HMXB) discovered so
far to emit the largest part of their luminosity at high energies
\citep{hermsen1977,gregory1978,albert2006}, being therefore a member
of the class of $\gamma$-ray binaries.  Variability of its emission,
at the timescale set by the $\sim$26.5 d orbital period, has been
found at almost all wavelengths, e.g.,
\citet{albert2008,abdo2009,torres2010,zhang2010}.  The companion star
is a massive B0Ve star, with a mass between 10 and 15 $M_{\odot}$, in
an eccentric 26.5 d orbit \citep{casares2005}. For the nature of the
compact object in $\gamma$-ray binaries, models involving an accreting
black hole launching a relativistic jet \citep[the microquasar
  scenario; see, e.g.][and references therein]{boschramon2009} and a
rotation-powered neutron star (NS in the following) emitting a
relativistic wind of particles \citep[see,
  e.g.][]{maraschi1981,dubus2006,sierpowska2008}, have been proposed.

The presence of a NS in {\lsi} would be definitely proven by the
observation of pulsations, but deep searches in the radio
\citep{mcswain2011} and X-ray band \citep{rea2010} were not
successful, so far. This is not surprising, since free-free absorption
easily washes out the pulses in the radio band, while the upper limit
of $\approx10\%$ (3 $\sigma$ confidence level) on the pulsed fraction
in X-rays could well be larger than the actual pulsed fraction of the
source. However, in the past few years, a couple of energetic
($\approx 10^{37}$ erg s$^{-1}$), short ($\simlt 0.3$ s) bursts were
detected by the {\it Swift}-Burst Alert Telescope from a region of a
few arc minutes of radius, compatible with the position of {\lsi}
\citep[][see Table~\ref{tab:bursts} for their observed
  properties]{depasquale2008,barthelmy2008,burrows2012}.  The
properties of the two bursts are typical of those observed in
magnetars, namely NSs for which emission is believed to be powered by
their strong magnetic energy.  It is probable that the bursts were
emitted by \lsi\ itself. Otherwise, we would be witnessing the
unlikely alignment, within a couple or arcmin, of a gamma-ray binary
(a population of objects for which a handful members are known) with a
magnetar-like burst-emitting object (for which we know 20 sources). If
the \lsi\ origin is accepted, any model of its multi-wavelength
emission should thus provide an explanation of such bursts.

Under the common assumption of pulsars emitting their rotational
energy via magnetic dipolar losses, the NS surface dipolar magnetic
field can be estimated from the observed period and period derivative
\citep{pacini1967,gold1969}. For the known magnetars, this usually
ranges from $\sim5\times10^{13}$ to $2\times10^{15}$\,G; recently,
however, two sources with a lower field, $\simgt 7\times10^{12}$ G,
were discovered, the emission of which is still believed to be powered
by non-dipolar components of the magnetic field
\citep{rea10,turolla2011,rea2012}.  About 20 magnetars are known to
date, all being isolated pulsars with periods ranging from 0.3--12\,s,
usually young spin-down ages ranging between 0.7--230 kyr (again with
the two exceptions reported above which are also much older systems),
and X-ray luminosities of the order of $10^{33-35}~$erg~s$^{-1}$ (see
\citealt{mereghetti2008,rea11} for recent reviews).  Magnetars,
historically divided into the two subclasses of Anomalous X-ray
Pulsars (AXPs) and the Soft Gamma Repeaters (SGRs), display a large
variety of bursts and flares, with properties at variance with those
observed from other compact objects such as accreting NSs or
BHs. Magnetars bursts can be empirically divided in three main classes
(although there is probably a continuum among them): the short bursts
($\sim0.01-1$\,s; $10^{37-40}~$erg~s$^{-1}$), the intermediate bursts
($\sim5-50$\,s; $10^{40-42}~$erg~s$^{-1}$) and the giant flares
($\sim100-500$\,s; $10^{43-47}~$erg~s$^{-1}$).

\citet{torres2012} have started to study how a high-field NS could
cope with the multi-wavelength phenomenology of {\lsi}.  In their
scenario the NS would behave as a usual rotation-powered pulsar only
when far from the companion star, whereas close to periastron, the
increased pressure exerted by the matter of the Be equatorial disk
would rather overcome the pulsar pressure, quenching the
rotation-powered pulsar behavior.  Though, accretion of the matter
captured would be inhibited by the quick rotation the NS, which would
then act as a propeller \citep{illarionov1975}. Such alternation
between ejector and propeller states along the orbit, which we refer
to as a {\it flip-flop} state, was originally proposed by
\citet{gnusareva1985} for NS in close-binary orbits of high
eccentricity, and already applied to the case of {\lsi} by
\citet{campana1995} and \citet{zamanov1995}.

In this paper we delve further into this scenario, estimating the
interval of spin periods at which a NS in {\lsi} is expected to behave
either as an ejector or a propeller, and the duration of each of
the states experienced by a NS in an eccentric binary system, as
it spins down during its initial evolution.  By taking into account
the constraints set on the parameters of the system by the observed
$\gamma$-ray luminosity, we also estimate the relative likelihood of
observing the assumed NS in one of the different states. Results
  are discussed, comparing the case of an assumed high-field NS in
{\lsi} to possibly related systems, such as rotation powered sources
in eccentric binary systems, as well as very long period HMXBs,
thought to have host a magnetar in their early evolutionary stages.

\begin{deluxetable}{lrr}
\tablecaption{Bursts observed by {\it Swift}-BAT from
  {\lsi}\label{tab:bursts}} \tablehead{ & \colhead{Burst
    No.\Rmnum{1}\tablenotemark{(a)}} & \colhead{Burst
    No.\Rmnum{2}\tablenotemark{(b)}} } \startdata Date & 2008 Sep 10
& 2012 Feb 5 \\ Position uncertainty & 2.1' & 3' \\ Ang. sep. & 0.60'
&1.07' \\ T$_{100}$ (s) & 0.31 & 0.044 \\ Fluence ($10^{-8}$ erg
cm$^{-2}$) & $1.4\pm0.6$ & $0.58\pm0.14$ \\ $\Gamma$ & $2.0\pm0.3$ &
$3.9\pm0.4$ \\ Luminosity ($10^{37}$ erg s$^{-1}$) & 2.1 & 6.3
\enddata \tablenotetext{(a)}{\citet{torres2012}}
\tablenotetext{(b)}{From \citet{burrows2012},  see also
  {http://gcn.gsfc.nasa.gov/notices\_s/513505/BA/}} \tablecomments{The
  positional uncertainty is given at a 90$\%$ confidence level,
  including also systematic uncertainties. The angular separation is
  calculated with respect to the position of the optical counterpart.
  The $T_{100}$ duration and the fluences are estimated in the 15--50
  keV band. Burst spectra were fitted by a power law with index
  $\Gamma$. The average luminosity is estimated by assuming a distance
  of 2 kpc \citep{frail1991}. }
\end{deluxetable}

\section{Spin evolution of a NS}
\label{sec:spin}

A NS evolves through different emission mechanisms during its
existence, ejector, propeller, accretor, and georotator, depending on
the balance between the outward pressure exerted by its
electromagnetic field, and the ram pressure of the surrounding matter
\citep[see, e.g.,][and references therein]{lipunov1992,
  ghosh2007}. The electromagnetic pressure critically depends on the
spin period of the NS, $P$, and on the strength of its dipolar
magnetic field, $B_1$. On the other hand, if the NS has a high mass
companion, the pressure exerted by the mass lost by the latter,
through a wind and possibly an equatorial disc such as in the case of
a Be star, is mainly determined by the density and velocity of the
outflow, and by the velocity of the NS motion along the orbit.  It
turns out that, once the NS magnetic field and the rate of mass
captured by the NS, $\dot{M}_1$, are set, the state in which the NS
lies is determined by its spin period, $P$. While at fast spin rates
the NS behaves as an ejector, it is expected to become a propeller
first, and subsequently accrete matter on its surface as it slows
down.

 If the NS orbit is highly eccentric, the rate of mass captured by the
 NS may vary by orders of magnitude along an orbital cycle, even if
 the companion star is assumed to lose mass at a constant rate; along
 its orbit the NS may then switch from one state to the other, such as
 it is the case for the {\it flip-flop}, ejector/propeller state
 proposed for {\lsi}.

\subsection{Ejector state}
\label{sec:ejector}

In the ejector state, a NS spinning at an angular frequency,
$\Omega=2\pi/P$, emits energy across the entire electromagnetic
spectrum at the expenses of its rotational
energy. \citet{spitkovsky2006} evaluated the spin-down luminosity of a
strongly magnetized oblique rotator, by solving for the dynamics of
the field in presence of conducting plasma (the so-called {\it
  force-free} limit of relativistic magneto-hydrodynamics):
\begin{equation}
L_{\rm ej}=(B_1
  R_1^3)^2\frac{\Omega^4}{c^3}(1+\sin^2{\alpha}).
\label{ejectorlum}
\end{equation} 
Here $B_1$ is the dipolar magnetic field at the equator of the
NS, $R_1$ is the NS radius and $\alpha$ is the angle between the
magnetic and the spin axis.  The spin-down torque acting on the NS can
be therefore expressed by
\begin{equation}
N_{\rm ej}=-\frac{L_{\rm ej}}{\Omega}=-(B_1
  R_1^3)^2\frac{\Omega^3}{c^3}(1+\sin^2{\alpha}).
\label{ejector}
\end{equation}

According to the conventional pulsar models (\citealt{goldreich1969};
see \citealt{lipunov1992} for a review), a NS behaves as an ejector as
long as it manages to keep the surrounding plasma from penetrating
into its light cylinder, the radius of which is $
R_{LC}={c}/{\Omega}.  $ To stop the in-fall of the matter captured by
the gravitation of the NS before it penetrates into the light
cylinder, the pressure exerted by the NS electromagnetic field must
overcome the pressure of the in-falling matter. Following
\citet{bondihoyle1944}, the radius at which matter is captured by the
gravitational field of the NS is:
\begin{equation}
\label{eq:gravradius}
R_G=\frac{2 G M_1}{v_{rel}^2},
\end{equation}
where $M_1$ is the NS mass and $v_{rel}$ is the velocity of the
captured matter with respect to the NS.  At lower radii, matter would
start falling towards the NS at a velocity of the order of the
free-fall value,
\begin{equation}
v_{ff}=\sqrt{\frac{2GM_1}{r}},
\end{equation} 
exerting a pressure
\begin{equation}
p_{\rm ram}\approx \rho v_{ff}^2 = \frac{\dot{M}_1}{4\pi}\frac{\sqrt{2GM_1}}{ r^{5/2}},
\end{equation}
where $\rho$ is the gas density, and the mass continuity equation was
used. The pressure of the NS electromagnetic field outside the light
cylinder,
\begin{equation}
p_{\rm ej}= \frac{L_{\rm ej}}{4\pi c r^2},
\end{equation} scales less steeply
with the distance than the pressure of the incoming matter inside the
gravitational radius ($p_{\rm ram}\propto r^{-5/2}$); when the
pressure of the captured matter evaluated at $R_G$ overcomes the
electromagnetic pressure, it is then expected to penetrate into the
light cylinder as it falls inwards, driving the NS out of the ejector
phase as a consequence. The matter in-fall may be stopped in fact only
by the NS magnetospheric pressure,
\begin{equation}
p_{\rm magn}=(B_1R_1^3)^2\frac{1}{8\pi r^6}.
\end{equation} 
In this case, the size of the magnetosphere is indeed defined in terms
of the balance between $p_{\rm magn}$ and $p_{\rm ram}$, yielding the
so-called Alfven (or magnetic) radius,
\begin{equation}
R_M = \frac{(B_1 R_1^3)^{4/7}}{\dot{M}^{2/7}(2GM_1)^{1/7}}.
\label{RM}
\end{equation} 

When the magnetosphere is able to extend up to the light-cylinder
radius again (e.g., because of a decrease of the mass capture rate),
the NS may then resume to emit as an ejector. The condition to recover
an unscathed light cylinder [$R_M \geq R_{LC}$, i.e., $p_{\rm
    ram}(R_{LC})\leq p_{\rm mag} (R_{LC})$] is slightly different than
the condition to stop the ejector mechanism [$p_{\rm ram}(R_G)\geq
  p_{\rm ej}(R_G)$]. The former condition is fulfilled at a larger NS
spin period when the other relevant magnitudes are held fixed (see,
e.g. the discussion in \citealt{torres2012}), and can be considered
more restrictive to identify when the NS abandons the ejector
state. For simplicity and to be conservative, we then consider
throughout this paper the equality, $R_M=R_{LC}$, to define the
transition either from, and into the ejector state.  The period at
which the transition from the ejector to the propeller state takes
place can be expressed in terms of the rate of mass captured by the
NS, once the the mass, radius and magnetic field of the NS are fixed:
\begin{equation}
P_{\rm ej\rightarrow
  sup\:prop}(\dot{m}_1)=0.24\:\:b_1^{4/7}\:m_1^{-1/7}\:r_1^{12/7}\:\dot{m}_1^{-2/7}\:{\rm
  s}.
\label{eq:p1}
\end{equation}
Here, $b_1=(B_1/10^{13}\: \rm G)$, $m_1=(M_1/1.4\: M_{\odot})$,
$r_1=(R_1/10 \:\rm km)$ and $\dot{m}_1=(\dot{M}_1/10^{17}$ g s$^{-1})$
are the magnetic field, the mass, the radius, and the rate of mass
captured by the NS, in units of the values we consider as {\it
  fiducial} in the rest of the paper (see Table~\ref{tab:scales} for a
complete list of the scale units considered).

\begin{deluxetable}{lll}
\tablecaption{Scale units used in this paper
\label{tab:scales}} 
\tablehead{ \colhead{Scale} & \colhead{Definition} & \colhead{} }
\startdata 
$b_1$ & $B_1/10^{13}\: G$ & NS magnetic field \\
$m_1$ & $M_1/1.4\: M_{\odot}$   & NS mass \\
$r_1$ & $R_1/10\: \rm km$ & NS radius \\
$\mathcal I$ & $I_1/10^{45}$ g cm$^2$ & NS moment of inertia \\
$\dot{m}_1$ & $\dot{M}_1/10^{17}$ g s$^{-1}$ & NS mass capture rate \\
$\dot{m}_1^{max}$ & $\dot{M_1}^{max}/10^{17}$ g s$^{-1}$& max NS mass capture rate \\
$\dot{m}_1^{min}$ & $\dot{M_1}^{min}/5\times10^{12}$ g s$^{-1}$ & min NS mass capture rate \\
$b_2$ & $B_2/0.6$ kG & Be star magnetic field \\
$m_2$ & $M_2/12.5\: M_{\odot}$ & Be star mass \\
$r_2$ & $R_2/10\: R_{\odot}$ & Be star radius \\
$n_2$ & $n/2$ & index of Be disc mass capture  \\
 & & rate radial dependence \\
$d_7$ & $d_{cut}/7 R_{2}$ & Be disc cut-off size \\
$\dot{m}_2^p$ & $\dot{M}_2^p/10^{18}$ g s$^{-1}$ & Be star mass loss rate \\
$v$ & $v_{\infty}^p/10^8$ cm s$^{-1}$ & Be star wind term. velocity 
\enddata
\end{deluxetable}

\subsection{Supersonic propeller state}
\label{sec:propeller}

When the NS stops acting as an ejector, accretion onto its surface is
still inhibited by the rotation of the NS.  The magnetosphere spins
much faster than the in-falling matter at the boundary defined by
$R_M$, and the interchange instabilities allowing the the plasma to
enter into the magnetosphere are strongly suppressed
\citep{elsner1977}.  To express the centrifugal inhibition of
accretion, we introduce the co-rotation radius, defined as the radius
at which the linear velocity of the rotating magnetosphere equals the
Keplerian rate, $\Omega_K(r)=(GM_1/r^3)^{1/2}$:
\begin{equation}
R_{co}=\left(\frac{GM_1}{\Omega^2}\right)^{1/3},
\end{equation}
and define the NS fastness as \citep{ghoshlamb1979III},
\begin{equation}
\omega_*=\frac{\Omega}{\Omega_K(R_M)}=\left(\frac{R_M}{R_{co}}\right)^{3/2}.
\end{equation}
If $R_M > R_{co}$ (i.e. $\omega_* > 1$), a centrifugal barrier
prevents the accretion of matter onto the NS surface, and the NS is
said to lie in a {\it propeller} state.  \citet{illarionov1975}
expressed the luminosity emitted by the NS in this state in terms of
the energy needed to balance gravitational energy of in-falling
matter, $L_{\rm prop}^{IS}= \dot{M} v_{ff}^2 / 2$. Subsequent studies
\citep{davies1979,davies1981,mineshige1991} showed how a quasi-static
corona forms around the NS, as far as the quickly rotating
magnetosphere transfers energy to the incoming matter at a rate larger
than the gas cooling rate. Such a corona extends down to the radius
where its pressure is balanced by the pressure of the magnetic
field\footnote{A proper assessment of the magnetospheric boundary
  depends on the details of the structure of the corona. However, the
  ratio between $R_{in}$ and $R_M$ evaluated by a number of authors
  (e.g. \citealt{davies1981} who evaluated it as $(R_G/R_M)^{2/9}$) is
  of the order of one for the parameters considered here. In light of
  the large uncertainties on the plasma capture process (see
  \S\ref{sec:mass}), we then consider $R_{in}=R_M$ to make simpler the
  evaluation of the torques.}, $R_{in}\approx R_M$.  At the interface
between the corona and the magnetosphere the gas is shocked by the
supersonic motion of the field lines, and energy is transferred to the
coronal gas through turbulent or convective motions \citep[see][for a
  detailed treatment]{wang1985}. Such a transfer takes place at the
expenses of the spin of the NS, which decelerates at a rate
\citep[see, e.g.,][]{mineshige1991}:\vspace{0.2cm}
\begin{equation}
N_{\rm prop}=\frac{L_{\rm prop}}{\Omega}\approx\frac{1}{\Omega}\:\epsilon \: \times \: 4\pi R_M^2 v_{\rm t}(R_M),
\label{superprop}\vspace{0.2cm}
\end{equation}
where $\epsilon$ is the energy density transferred by the NS during
each revolution, $v_{\rm t}$ is the velocity of the turbulence
developing at $R_M$, and $4\pi R_M^2 v_t(R_M)$ is volume of the gas
perturbed by the motion of the dipolar magnetic field, assumed to be
tilted with respect to the spin axis.  As long as the linear velocity
of the magnetosphere exceeds the sound speed at $R_M$ (taken to be of
the order of the free-fall velocity, $v_{ff}$), and the energy
  released to the corona by the NS dominates radiative losses
  \citep[see the discussion of][relative to wind-fed close binary
    systems]{ikhsanov2002}, the propeller is considered {\it
  supersonic} and the turbulent motions take place at a velocity
$v_{\rm t}\simeq v_{ff}$.  Though, there is no general consensus on
the estimate of the propeller efficiency. \citet{davies1979} and
\citet{davies1981} consider $\epsilon\simeq\rho v_{ff}^2 /2$,
recovering the scaling of \citet{illarionov1975}
\begin{equation}
N_{\rm prop}^{IS} = -\dot{M}\sqrt{GM_1 R_M}\;\omega_*^{-1}.
\label{is}
\end{equation} 
On the other hand, if $\epsilon\simeq \rho(\Omega_S R_M)^2/2$ is
considered, a much stronger torque is obtained
\citep{mineshige1991,ghosh1995a},
\begin{equation}
N_{\rm prop}^{G}  =  -\frac{1}{6}\dot{M}\sqrt{GM_1 R_M}\;\omega_*,
\label{ghosh}
\end{equation} 
where the numerical factor takes into account the degree of
non-axisymmetry of the magnetosphere with respect to the spin axis
\citep[see also][who derived expressions with a similar
  scaling]{wang1985,illarionov1990,bisnovatyikogan1991}.  The
difference between the two estimates given by Eq.~(\ref{is}) and
(\ref{ghosh}) may be as large as $\approx 10^4$, when a NS with the
fiducial parameters defined in Table~\ref{tab:scales} is rotating
close to the critical period marking the transition between the
ejector and the propeller state (Eq.~\ref{eq:p1}), since
$\omega_*\approx 100$ in that case.  The discrepancy between the two
propeller torques yields a significant uncertainty in the evaluation
of the timescale of the NS evolution in the propeller state (see,
e.g., the discussion of \citealt{francischelli2002} and
\citealt{mori2003}). In order to be conservative when describing the
evolution of {\lsi} in the {\it flip-flop} state, we consider the two
estimates given above as limiting cases.

\subsection{Subsonic propeller and onset of accretion}

As the velocity of the NS decreases and becomes comparable to the
speed of sound at the inner boundary of the corona, the propeller
becomes {\it subsonic}, with the turbulence traveling at a speed set
by the NS rotation, $v_{\rm t}\simeq\Omega R_M$ \citep{davies1981}. We
assume that such a transition takes place when $ R_M\simeq R_{co} $,
translating into a period:
\begin{equation}
P_{\rm sup\:prop\rightarrow
  sub\:prop}=18\:\:b_1^{6/7}\:\dot{m}_1^{-3/7}\:m_1^{-5/7}\:r_1^{18/7}\:\mbox{s}.
\label{eq:p2}
\end{equation} 
 From Eq.~(\ref{superprop}) it is deduced how the torque experienced
 by the NS in this stage differs by a factor $\Omega R_M/v_{ff}(R_M)$
 from those defined in the previous section. The rate at which energy
 is transferred from the NS to the surrounding corona decreases with
 increasing period; gas in the corona then starts to cool down,
 facilitating the plasma entry into the magnetosphere. A fraction of
 the incoming matter may then accrete down to the NS surface already
 in the subsonic propeller state (the so-called settling regime
 studied by \citealt{shakura2012}).  Subsequently, the cessation of
 any significant barrier effect is achieved when the energy released
 by the rotating magnetosphere to the incoming matter can be neglected
 with respect to the cooling of the gas; the spin period for such a
 transition was estimated as:
\begin{equation}
P_{\rm sub\:prop\rightarrow
  acc}=91\:\:b_1^{16/21}\:\dot{m}_1^{-5/7}\:m_1^{-4/21}\:r_1^{16/7}\:\mbox{s},
\label{eq:pmax}
\end{equation}
by \citet{ikhsanov2001}, who used this expression to estimate the
duration of the subsonic propeller state in wind-fed binary systems
\citep[see also][]{ikhsanov2007}.

\section{Magnetic dissipative torques}
\label{sec:magn}

The interaction between the strong magnetic field of a magnetar and
the field of a low mass, convective companion star was invoked by
\citet{pizzolato2008} to suggest how the spin period of the X-ray
source 1E 161348--5055, could have been locked to the orbital period
of the system, similar to what happens in polar cataclysmic variables
\citep[see e.g.][]{warner1995}.  In such a case, it is in fact argued
that a dissipative torque,
\begin{equation}
N_{\rm magn}\approx\frac{\mu_1\mu_2}{d^3},
\label{eq:magn}
\end{equation}
 develops as the magnetic dipole of the white dwarf, $\mu_1$, and the
 magnetic field of the companion star, $\mu_2$, interact at an orbital
 separation $d$.  The magnetic field of the companion star can be
 either induced by the white dwarf field
 \citep[e.g.][]{joss1979,lamb1983,campbell1984} or intrinsic to the
 companion star \citep{campbell1985,hameury1987}. The rotation of a
 low mass companion star belonging to a close system ($P_{orb}\approx$
 few hours) is synchronized to the orbital motion by tidal forces on a
 relatively short timescale, $\approx 10^2-10^3$ yr, as it is obtained
 by considering the relation given by \citet{zahn1977}
\begin{equation}
t_{sync}=\frac{1}{6}\left(\frac{M_2}{M_1}\right)^2\left( \frac{M_2 R_2^2}{L_2} \right)^{1/3} \frac{I_2}{k_2 M_2 R_2^2} \left( \frac{a}{R_2}\right)^6.
\label{eq:synchro}
\end{equation}
Here, $M_2$, $R_2$, $L_2$ and $I_2$ are the mass, radius, luminosity
and momentum of inertia of the non degenerate star, $k_2$ is the
constant of apsidal motion and is of the same order of $I_2/M_2 R_2^2$
\citep{zahn1977}, and $a$ is the semi-major axis of the orbit.  The
torque due to the interaction between the magnetic fields of the two
stars then acts to bring the white dwarf to synchronicity with the
orbit, as well.

The projected rotational velocity of the $\sim 12.5$ $M_{\odot}$ Be
star in {\lsi} was measured by \citet{casares2005} as $113$ km
s$^{-1}$. Such a value corresponds to a spin period of $\simeq
4.5\:r_2\:\sin{i}$ d, where $i$ is the inclination of the system and
$r_2=(R_2/10 R_{\odot})$ is the radius of the companion in units of 10
$R_{\odot}$. Plainly put, the rotation of the Be star is not locked to
the orbital period of the system. This is consistent with the time
needed to synchronize the spin of the non-degenerate star to the orbit
of {\lsi} through tidal interactions with the compact object ($\approx
10^7$ yr , see Eq.~(\ref{eq:synchro})).

However, the magnetic field of a star with the properties of the Be
star in {\lsi} can be in principle very intense. A surface field of
$\sim$0.6 kG was measured from a star of the same luminosity class
\citep{petit2008}, while even larger fields were measured from
peculiar stars with spin periods lower than 1 d \citep[see
  e.g. table~1 in][and references therein]{oskinova2011}.  An
intrinsic field of this order is larger by orders of magnitude than
any field which may be induced by the NS field, and the magnetic
dipole moment of the Be star, $\approx 10^{38}$ G cm$^{-3}$, would be
much larger than the NS moment, $\approx 10^{31}$ G cm$^{-3}$, as
well. However, in the case of a relatively wide binary such as {\lsi},
the steep dependence of Eq.~(\ref{eq:magn}) on the orbital separation
greatly reduces the magnitude of the torque.

\section{Mass capture}
\label{sec:mass}

The rate of mass lost by a Be star such as that in {\lsi},
$\dot{M}_2$, is given by the sum of the mass lost through a {\it fast}
polar wind, $\dot{M}_2^p$, and a {\it slow} equatorial disc,
$\dot{M}_2^d$ \citep[e.g.][]{waters1988}. According to the Bondi-Hoyle
description, the mass lost by the companion star is then captured by
the NS at a rate, $\dot{M}_1=\dot{M}_2\:(R_G/d)^2/4$, where $R_G$ is
the radius of gravitational capture defined by
Eq.~(\ref{eq:gravradius}). The estimate of $\dot{M}_1$ obtained under
this approximation is thus very steeply dependent on the velocity of
the captured mass with respect to the NS, $v_{rel}$. The velocity of
the polar, radiatively driven outflow is described by $v^p(r)\simeq
v_{\infty}^p (1-R_2/d)$, where $v_{\infty}^p$ is the polar wind
terminal velocity \citep[e.g.][]{lamers1999}. The orbital velocity of
the compact object in {\lsi} can be neglected with respect to the wind
velocity \citep[see Fig.~13 of][]{torres2012}, and the rate of mass
captured from the polar wind is given by:
\begin{equation}
\dot{M}_1^p=\dot{M}_2^p \frac{(GM_1)^2}{(v_{\infty}^p)^4
  d^2}\left(1-\frac{R_2}{d}\right)^{-4}.
\label{mdotwind}
\end{equation}
In the following we scale the rate of mass lost by the Be star and the
wind terminal velocity in units of $\dot{m}_2^p=(\dot{M}_2^p/10^{18})$ g
s$^{-1}$ and $v=(v_{\infty}^p/10^8$ cm s$^{-1})$, respectively, of the
order of the  estimates given by \citet{waters1988}.

A rapidly rotating Be star loses mass through an equatorial disc at a
rate which is generally $10-100$ times larger than that of the polar
wind \citep[see, e.g.,][]{lamers1987}. By modeling the observed IR
excess, \citet{marti1995} estimated the equatorial disk mass loss rate
of the Be star in {\lsi}, $M_2^d$, to lie between 0.25 and
2.5$\times10^{19}$ g/s. This range depends on the value of the radial
velocity of the flow at the surface of the Be star, assumed to vary in
the range 2--20 km s$^{-1}$. However, modeling the capture of mass
from the equatorial disc is much more uncertain with respect to the
polar wind case.  In this case, for any reasonable assumption on the
radial and azimuthal velocity profile of the equatorial disk matter,
the orbital velocity of the compact object in {\lsi} cannot be
ignored, as it is the case when the poloidal wind is considered. Most
importantly, the Bondi-Hoyle approximation yields values of the mass
capture rate at the periastron of the orbit which may be unphysical
larger than the rate of mass lost by the Be star, essentially because
of the low relative velocity of the disc matter in such a system
($\simlt 10^7$ cm s$^{-1}$; see Fig.~13 in
\citealt{torres2012}). Given these large uncertainties, to model the
capture of mass from the equatorial disc we parametrize with a power
law its dependence on the orbital separation, $d$, adding a strong
cut-off at a distance $d_{cut}$ to reproduce a tidal truncation of the
disc due to the interaction with the compact object
\citep{okazaki2002}. We thus consider:
\begin{equation}
\dot{M}_1^d=\dot{M}_1^{max}\left(\frac{d}{d_{min}} \right)^{-n}\:\exp{\left[-\left(\frac{d}{d_{cut}}\right)^{10}\right]}.
\label{mdotdisc}
\end{equation}
Here, $\dot{M}_1^{max}=\dot{M}_1^d(d_{min})$ is the maximum rate of
mass captured by the NS from the equatorial disc, when the NS is close
to the periastron of the orbit, i.e., at an orbital separation
$d_{min}=a(1-e)$. In the following, we consider the eccentricity
$e=0.63\pm0.11$ measured by \citet{casares2005}, and compatible with
the estimates of \citet{grundstrom2007} and \citet{aragona2009}, while
the semi-major axis of the orbit, $a$, is estimated as
$6.3\times10^{12}$ cm by using the third Kepler law for a system with
an orbital period of 26.5 d and a total mass of 14 M$_{\odot}$.  Since
the disc is much denser than the polar wind, the maximum rate of mass
captured by the NS, $\dot{M}_1^{max}$ is equal the rate of mass
captured from the disc, $\dot{M}_1^{d,max}$; in the following we scale
this value in units of $\dot{m}_1^{\rm max}=\dot{M}_1^{max}/10^{17}$ g
s$^{-1}$, which was also used by \citet{dubus2006}. Such a value is
roughly in between the estimates considered by \citet{zamanov1995} and
\citet{gregoryneish2002}, $\approx3\times10^{17}$ and
$\approx0.6\times10^{17}$ g s$^{-1}$, respectively. A value of the
same order, $\sim 0.5\times10^{17}$ g s$^{-1}$, was also found by
\citet{romero2007} from simulations of the interaction between the
equatorial disk of the Be star in {\lsi} and the compact object,
assumed to be accreting in the case they considered.  It is also worth
to note how a peak accretion rates up to $\sim 6\times10^{17}$ g
s$^{-1}$, is deduced from the X-ray luminosity observed at the peak of
the outburst shown by the pulsar, 4U 0115+63 \citep[see][and
  references therein]{ferrigno2011}, which has a B0.2Ve star companion
and orbital parameters ($P_{orb}=24.3$ d, $e=0.34$) similar to those
of {\lsi}. The index of the power law $n$ of Eq.~(\ref{mdotdisc}) is
varied between 1 and 3, to qualitatively reproduce the dependence of
the mass capture rate on the distance found by the simulations of
\citet{romero2007}, and in particular the ratio in the range
$10$--$100$ between the maximum and minimum mass capture rate in the
absence of a significant cut-off \citep[see the solid curves in the
  left panel of Fig.~13 plotted by ][]{torres2012}. The fiducial unit
of the truncation radius of the equatorial disc is set to $7$\,R$_2$,
of the order of the estimates given by \citet{grundstrom2007} on the
basis of the observed equivalent width of the H$\alpha$ emission
line. Larger values are also reported in literature \citep[see
  e.g.,][who give $d_{cut}\sim12$\,R$_2$]{gregoryneish2002}.

\begin{figure}
\epsscale{1.2} \plotone{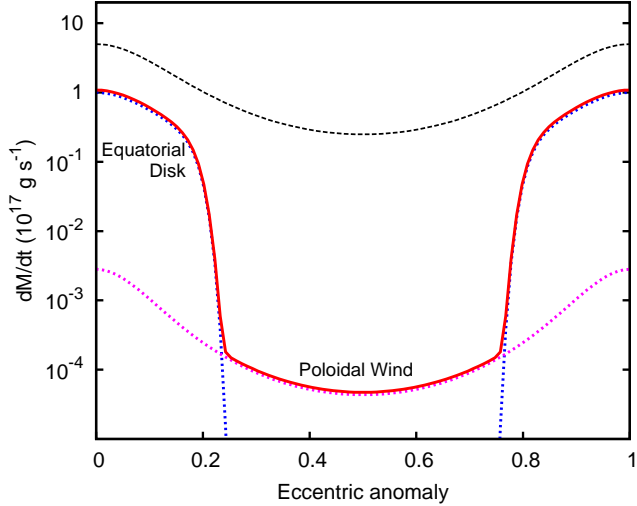}
\caption{Rate of mass captured by a NS from the equatorial disc (blue
  dotted line) and the poloidal wind (magenta dotted line) emitted by
  the Be star in {\lsi}, as a function of the eccentric anomaly, and
  for the fiducial values of the relevant parameters (see
  Table~\ref{tab:scales}). Red solid line is the sum of these two
  contributions. The black dashed line shows the case of an increased
  Be star mass loss rate, $\dot{m}_1^{max}=5$, and a disc cut-off
  beyond the maximum orbital separation, $d_{cut}>
  a(1+e)$. \label{fig:mdot}}
\end{figure}

 To evaluate the orbital dependence of the wind and disc contributions
 to the total rate of mass captured by the compact object,
 $\dot{M}_1=\dot{M}_1^d+\dot{M}_1^p$, we use the relation
 $d=a(1-e\cos{\epsilon})$ to express the orbital separation in terms
 of the eccentric anomaly, $\epsilon$. A red solid line shows in
 Fig.~\ref{fig:mdot} the mass capture rate as a function of
 $\epsilon$,  for the fiducial set of parameters.  In the following
 we shall also consider a mass capture rate increased by a factor of
 five, with an outer radius exceeding the maximum orbital separation
 (see black dashed line in Fig.~\ref{fig:mdot}), in order to mimic an
 enhancement of the rate of mass loss of the order of that reported by
 \citealt{gregory1989} and \citealt{zamanov1999} to explain the
 observed super-orbital variability (see e.g., \citealt{gregory2002}),
 with the possible expulsion of mass shells in the disc \citep[see,
   e.g.][]{gregoryneish2002}.

 \section{Timescales and changes of state}
\label{sec:model}

\begin{figure}[t!]
\epsscale{1.2}
\plotone{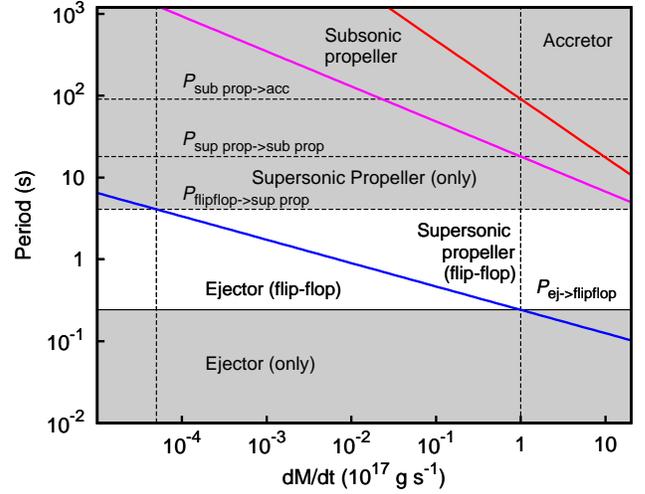}
\caption{States accessible to a system like {\lsi} in the NS spin
  period vs. mass capture rate phase space. Solid lines mark the
  transitions between the ejector, supersonic propeller, subsonic
  propeller, and accretor states, as defined in text [see equations
    (\ref{eq:p1}), (\ref{eq:p2}), and (\ref{eq:pmax})], and evaluated
  for $b_1=1$.  Vertical dashed lines mark the minimum and maximum
  capture rate experienced by the NS along its orbit, while the
  horizontal dashed lines indicate the values of the spin period at
  which transitions take place, when the relevant parameters are set
  equal to their fiducial values (see
  Table~\ref{tab:scales}).\label{fig:regimes}}
\end{figure}

The time it takes for a NS to reach a period $\bar{P}$, under the
action of a spin-down torque, $N(P)$, is obtained from the integration
of the equation,
\begin{equation}
N(P)=I\dot{\Omega}=-\frac{2\pi I}{P^2} \frac{dP}{dt},
\end{equation}
between the initial period, $P_0$, and $\bar{P}$:
\begin{equation}
\label{eq:timescale}
t=-2\pi I \int_{P_0}^{\bar{P}}\frac{dP}{P^2 N(P)}.
\end{equation}
A NS is generally considered to spin at its birth at a period of few
tens of ms, therefore spinning down as a rotation-powered pulsar. If
it belongs to a binary system such as {\lsi}, the ejector phase will
end as the pressure of the mass captured at periastron overcomes the
pulsar pressure. The system then enters in the {\it flip-flop} state
(i.e., the state at which it is in ejector phase in apastron and in
supersonic propeller in periastron) when its period attains a value
$P_{\rm ej\rightarrow flipflop}$. This is obtained by evaluating
Eq.~(\ref{eq:p1}) (plotted as a blue solid line in
Fig.~\ref{fig:regimes}, where the periods at which the various state
transitions take place are plotted as a function of the mass capture
rate, for a magnetic field $b_1=1$), at the maximum accretion rate
experienced by the NS along its orbit, $\dot{m}_{1}^{max}$ (rightmost
vertical dashed line in Fig.~\ref{fig:regimes}):
\begin{equation}
P_{\rm ej\rightarrow
  flipflop}=0.24\:\:b_1^{4/7}\:m_1^{-1/7}\:r_1^{12/7}\:(\dot{m}_{1}^{max})^{-2/7}\:{\rm
  s}.
\label{eq:Plow}
\end{equation}
Assuming for the moment the magnetic field of the NS to be constant,
the duration of the ejector phase is found by evaluating
Eq.~(\ref{eq:timescale}) between $P_0$ and $P_{\rm ej\rightarrow
  flipflop}$, with the ejector torque being described by
Eq.~(\ref{ejector}):
\begin{eqnarray}
t_{ej} & = &
6.3\;b_1^{-6/7}(\dot{m}_{1}^{max})^{-4/7}m^{-2/7}r^{-18/7}\mathcal{I}(1+\sin^2{\alpha})^{-1}\;\mbox{kyr},\nonumber
\\ & &
\label{eq:emtime}
\end{eqnarray}
where $\mathcal{I}=(I_1/10^{45})$ g cm$^2$ and $I_1$ is the NS moment
of inertia. The value of $t_{ej}$ does not significantly depend on
$P_0$ as long as it is much smaller than $P_{\rm ej\rightarrow
  flipflop}$.

\begin{figure}[t!]
\epsscale{1.2}
\plotone{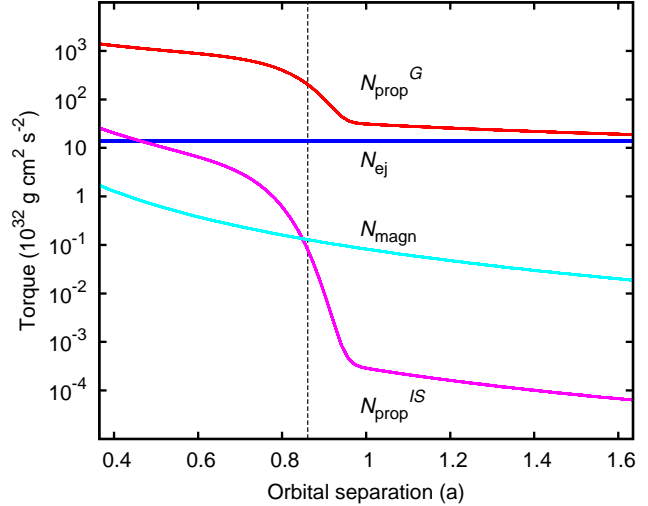}
\caption{ Propeller torque evaluated according to the two cases
  presented in text (red and magenta solid curves, respectively, see
  Eq.~[\ref{is}] and [\ref{ghosh}]), ejector torque (blue; see
  Eq.~[\ref{ejector}]) and magnetic torque (cyan; see
  Eq.~[\ref{eq:magn}]), plotted as a function of the orbital
  separation between the two stars of {\lsi} (expressed in units of
  the semi-major axis), evaluated for values of the relevant
  parameters set equal to the fiducial units (see
  Table~\ref{tab:scales}), and for an example period of $P=1$
  s. Vertical dashed line mark the critical separation,
  $d_{crit}(P=1{\rm s})$, at which the system performs a transition
  from ejector ($d>d_{crit}(P=1{\rm s})$) to propeller
  ($d<d_{crit}(P=1{\rm s})$).\label{fig:torques} }
\end{figure}

 Evaluated at a given period, Eq.~(\ref{eq:p1}) implicitly defines a
 critical value for the mass capture rate at which the NS switches
 from the ejector state to the propeller state (and vice versa),
 $\dot{M}_{crit}(P)$. By using the relations given in the previous
 section to express $\dot{M}_1$ as a function of the orbital
 separation, this condition translates into a critical separation
 $d_{crit}(P)$, defining the portion of the orbit during which the NS
 behaves as a propeller [for $d\leq d_{crit}(P)$], or as an ejector
 [for $d>d_{crit}(P)$].  We then express the torque experienced by the
 NS during the {\it flip-flop} state as:
\begin{displaymath}
N_{\rm ff}(d) = \left\{
\begin{array}{ll}
N_{ej} + N_{magn} & \textrm{for } d>d_{crit}(P)\\
N_{prop} + N_{magn} & \textrm{for } d\leq d_{crit}(P)
\end{array} \right.
\end{displaymath}
We summarize in Table \ref{tab} the spin evolution timescales,
$\tau=\Omega/\dot{\Omega}$, implied by the torques introduced in the
previous sections. The average torque experienced by the NS along one
orbit is: {\setlength\arraycolsep{2pt}
\begin{eqnarray}
<N_{ff}>&=&\frac{1}{P_{orb}}\int_{0}^{P_{orb}} N_{\rm ff}(t) dt ={}\nonumber\\
{}& & \frac{1}{P_{orb}} \int_{0}^{2\pi}   N_{\rm ff}[d({\mathcal M})] \left|\frac{dt}{d{\mathcal M}} \right| d{\mathcal M}.
\label{avtorque}
\end{eqnarray}} 
Here, ${\mathcal M}$ is mean anomaly, $|dt/d{\mathcal
  M}|=P_{orb}/2\pi$, the relation between the mean and the eccentric
anomaly is given by the Kepler's equation, ${\mathcal
  M}=\epsilon-e\sin{\epsilon}$, while the orbital separation is
obtained as $d=a(1-e\cos{\epsilon})$. The magnitude and dependence on
the orbital separation of the different torques introduced so far, as
well as the critical distance at which the transition
ejector/propeller takes place, evaluated for example at a period $P=1$
s and for the fiducial parameters defined in Table~\ref{tab:scales},
are plotted in Fig.~\ref{fig:torques}.

The NS abandons the {\it flip-flop} state when it reaches a period,
$P_{\rm flipflop\rightarrow sup\:prop}$, such that it is in a
propeller state at all points in the orbit, even when the rate of mass
capture is minimum, i.e., at apastron,
$\dot{M}_{crit}\simeq\dot{M}_1(d_{max})$. It can be seen from
Fig.~\ref{fig:mdot} that, as long as the equatorial disc is cut off at
a distance $d_{cut}<d_{max}$, this value is set by the rate of mass
captured by the polar wind, $\dot{M}_1^{p}(d_{max})$. Considering a
scale unit of $\dot{m}_1^{min}=\dot{M}_1^{min}/5\times10^{12}$ g
s$^{-1}$ for the minimum rate of mass capture, of the order of that
expected at the apastron of the orbit if the Be star loses mass
through the polar wind at a rate of $10^{18}$ g s$^{-1}$, the value of
$P_{\rm flipflop\rightarrow sup\:prop}$ is found from
Eq.~(\ref{eq:p1}):
\begin{equation}
P_{\rm flipflop\rightarrow sup\:prop}=4.1\:\:b_1^{4/7}\:
(\dot{m}_1^{min})^{-2/7}\:m_1^{-1/7}\:r_{1}^{12/7}\:\mbox{s}.
\label{eq:Phi}
\end{equation}  
Fig.~\ref{fig:regimes} shows how for the typical values of the
parameters relevant to the case of {\lsi}, the NS leaves the {\it
  flip-flop} state well before its propeller torque becomes subsonic,
and only the torques defined in $\S$~\ref{sec:ejector} and
$\S$~\ref{sec:propeller} are relevant to the evaluation of the time
spent by the NS in the {\it flip-flop} state.

\begin{deluxetable}{ll}
\tablecaption{Spin evolution timescales\label{tab}} \tablehead{
  \colhead{Torque} & \colhead{Timescale, $\Omega/\dot{\Omega}$ (kyr)}
} \startdata EM & 144$\:\:b_1^{-2}\:r_1^{-6}\:\mathcal{I}\:P^2$ \\ IS
Supersonic Propeller&
77.54$\:b_1^{4/7}\:\dot{m}_{1}^{-9/7}\:m_1^{-8/7}\:r_{1}^{12/7}\:\mathcal{I}\:P^{-2}$
\\ G Supersonic Propeller&
1.43$\:\:\:\:b_{1}^{-8/7}\:\dot{m}_{1}^{-3/7}\:m_1^{2/7}\:r_{1}^{-24/7}\:\mathcal{I}$\\ IS
Subsonic Propeller &
6.1$\:\:b_{1}^{-2/7}\:\dot{m}_{1}^{-6/7}\:m_{1}^{-3/7}\:r_{1}^{-6/7}\:\mathcal{I}\:P^{-1}$\\ G
Subsonic Propeller &
0.11$\:\:b_{1}^{-2}\:m_{1}\:r_{1}^{-6}\:\mathcal{I}\:P$\\ Magnetic
Torque &
$21\times10^3\:b_{1}\:r_{1}^{3}\:b_{2}\:r_{2}^3\:d_6^{-3}\:\mathcal{I}\:P^{-1}$ \enddata
\tablecomments{See Table~\ref{tab:scales} for the definition of the
  scale units. A value of $\alpha=45^{\circ}$ was considered to
  estimate the electromagnetic torque. $P$ is the NS spin period in
  seconds.}
\end{deluxetable}

\section{Results and constraints}

\subsection{The flip-flop phase}

The total time spent in the {\it flip-flop} phase by a NS in {\lsi},
is evaluated by integrating Eq.~(\ref{eq:timescale}) between $P_{\rm
  ej\rightarrow flipflop}$ and $P_{\rm flipflop\rightarrow
  sup\:prop}$. For the set of fiducial values we considered, the
source stays in a {\it flip-flop} state when its spin period lies in
the range between 0.24 and 4.1 s. The total time spent in this state
crucially depends on the relation considered to express the propeller
torque. Values of 25 and 282 kyr are obtained when the torque of
Eq.~(\ref{ghosh}) and (\ref{is}) are considered, respectively.  Such a
large discrepancy is due to the different dependence on the system
fastness $\omega_*$ of the two expressions. On the other hand, the
torque due to the interaction between the magnetic fields of the NS
and of the companion star (see \S~\ref{sec:magn}) has a negligible
impact on the evaluation of the {\it flip-flop} time-scale even if the
weakest propeller torque is considered.  The timescales obtained have
to be compared with the interval of $\approx 10^3$ kyr it would take
for a NS with the assumed fiducial values to cover the same interval
of periods, only by spinning down as a rotation-powered pulsar with a
constant magnetic field.

\begin{figure}[t!]
\epsscale{1.2}
\plotone{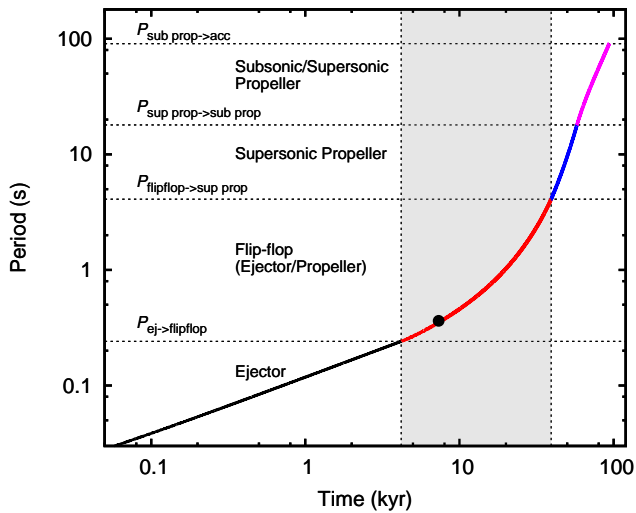}
\caption{Evolution of the spin period of a NS in a system such as
  {\lsi}, obtained considering the fiducial values of the system
  parameter (see Table~\ref{tab:scales}). Dashed horizontal lines mark
  the transition among different states, while the gray shaded area
  mark the time interval during which the NS is expected to lie in the
  {\it flip-flop} state. The black filled circle marks the spin period at
  which the system emits an ejector luminosity of $5\times10^{35}$ erg
  s$^{-1}$.
\label{fig:periodevol}}
\end{figure}

In the following, we focus on the results obtained with the stronger
propeller torque, Eq.~(\ref{ghosh}), since it gives values which can
be considered as conservative lower limits on the total time that a NS
in a system like {\lsi} is expected to spend in the {\it flip-flop}
phase.  We plot in Fig.~\ref{fig:periodevol} the evolutionary
track of the NS spin obtained by considering the fiducial parameters
defined above. The dependence of the total time spent by the system in
the {\it flip-flop} state can be approximated as
\begin{equation}
t_{\rm ff} \simeq 25 \: b_1^{-1.1} \: (\dot{m}_1^{max}/n_2)^{-0.3} \: d_7^{-1.1} 
\:(\dot{m}_1^{min})^{-0.12} {\rm kyr}.
\label{eq:flipfloptime}
\end{equation}
Here, $d_7=(d_{cut}/7 R_{2})$ and $n_2=(n/2)$. The {\it flip-flop} timescale
depends strongly on the strength of the NS magnetic field, and on the
amount of mass captured by the equatorial disc of the Be star, as it
is expressed by the dependence on its size, $d_{cut}$, and less
strongly on the maximum mass capture rate, $\dot{m}_1^{max}$.

Values of the magnetic field in excess of $ 2\times 10^{14}$ G reduce
the total {\it flip-flop} timescale to less than a kyr. Both, the
range of periods for which the NS is in the {\it flip-flop} state and
the magnitude of the spin down torque, increase when a stronger NS
magnetic field is considered, but the latter to a larger extent.  The
total {\it flip-flop} timescale also depends significantly on the
amount of mass captured by the equatorial disk of the Be star.  Such a
time varies between $\approx$50 and 15 kyr when the maximum rate of
mass captured varies between $10^{16}$ and $5\times10^{17}$ g
s$^{-1}$; a range of 17--36 kyr is obtained when the size of the disc
cut-off takes value between 10 and 5 $R_2$, while a smaller variation
of $\approx 15\%$ is obtained when values of $n$ in the range 1--3 are
considered.

\subsection{$B$-field evolution}

\begin{figure}[t!]
\epsscale{1.2}
\plotone{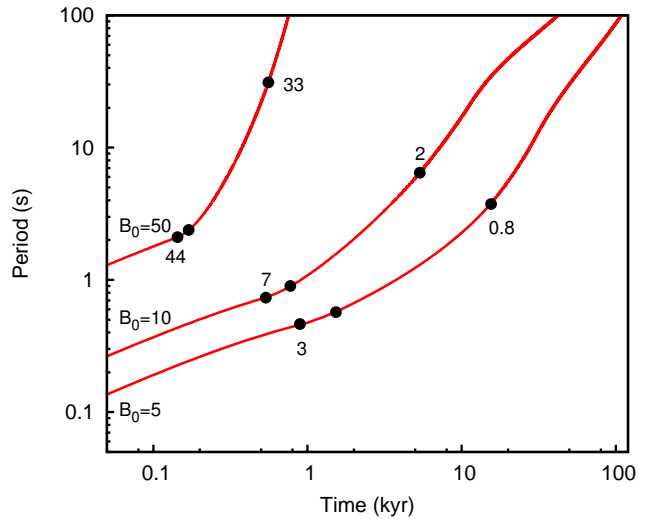}
\caption{Evolution of the spin period as a function of the time
  elapsed since the NS birth with $P_0=0.01$ s, for values of the
  initial NS magnetic field equal to (from top to bottom) $50$, $10$ and
  $5\times10^{13}$ G. From left to right along each track, black
  points refer to the ingress in the the {\it flip-flop} state, the
  period at which the system emits an ejector luminosity of
  $5\times10^{35}$ erg s $^{-1}$, and the egress from the {\it
    flip-flop} state. Numbers along each track denote the strength of
  the magnetic field at the ingress and egress from the {\it
    flip-flop} state, respectively, in units of $10^{13}$ G.
\label{fig:Bevol}}
\end{figure}

We have also studied the effect of the decay of the dipolar
magnetic-field on the total time spent by the system in the {\it
  flip-flop} state, by considering the simple relation given by
\citet{aguilera2008}
\begin{equation}
B_1(t)=B_0\frac{\exp{-t/\tau_O}}{1+(\tau_O/\tau_H)[1-\exp{(-t/\tau_O)}]}+B_{\rm as},
\end{equation}
where $\tau_H$ and $\tau_O$ are the timescales for Hall and Ohm decay,
set equal to $1$ and $10^3$ kyr, respectively, and $B_0$ and $B_{\rm
  as}$ are the initial and asymptotic value of the magnetic field,
respectively. The evolutionary tracks evaluated for $B_{\rm
  as}=5\times10^{12}$ G and a number of values of $B_0$ are plotted in
Fig.~\ref{fig:Bevol}. For the larger initial field values we
considered ($50\times10^{13}$ G), the time it takes for the NS to
enter the {\it flip-flop} state is so short ($\simeq 0.2$ kyr) with respect
to the assumed value of $\tau_H$, that no significant field decay has
still taken place and the timescale of the {\it flip-flop} state are
correspondingly very short ($\simlt$ kyr). On the other hand, values
in excess than 10 kyr are spent by the NS in the {\it flip-flop} state if
the initial field is $\simlt 5\times10^{13}$ G.

\subsection{Constraints from the apastron luminosity and Be-star mass loss rate variations}

\begin{figure}[t!]
\epsscale{1.2} \plotone{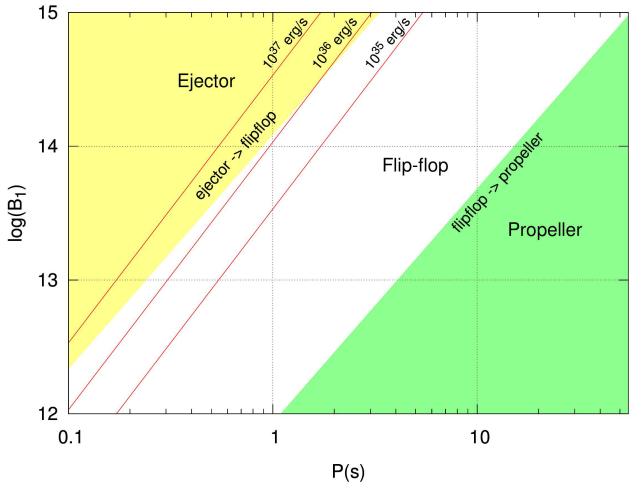}
\caption{ Ejector, {\it flip-flop} and propeller states plotted in the
  NS magnetic field vs. spin period phase space, evaluated for a NS in
  {\lsi} and for the fiducial values adopted for the maximum and
  minimum mass capture rate ($\dot{m}^{max}=1$, $\dot{m}^{min}=1$).
  From top to bottom, the red solid lines mark the relation between
  the period and the magnetic field of the NS when the ejector
  luminosity is $10^{37}$, $10^{36}$ and $10^{35}$ erg s$^{-1}$,
  respectively, and the magnetic offset angle is
  $\alpha=45^{\circ}$. \label{fig:BPregimes}}
\end{figure}

We plot in Fig.~\ref{fig:BPregimes} the different states (ejector,
{\it flip-flop}, propeller) in which a NS in {\lsi} is expected to lie
during the early stage if its evolution, depending on the values of
its magnetic field and spin period; the tracks delimiting the various
regions were evaluated from Eq.~(\ref{eq:Plow}) and (\ref{eq:Phi}),
for fiducial units of the maximum and minimum mass capture rate.

A relevant constraint on the possible position of {\lsi} in such
magnetic field vs. period diagram can be drawn on the basis of the
flux observed when the NS is close to apastron. According to the
simplest {\it flip-flop} scenario discussed \citep[i.e., not
  considering the possible effect of a relativistic wind on the
  spin-down of a magnetar, discussed by][]{harding1999}, when in the
apastron region the NS is powered by the ejector mechanism, and its
luminosity cannot exceed the value given by Eq.~(\ref{ejector}). By
evaluating such a relation for $\alpha=45^{\circ}$, and an ejector
luminosity of $10^{37}$, $10^{36}$, and $10^{35}$ erg s$^{-1}$, the
tracks plotted as red solid lines in Fig.~\ref{fig:BPregimes} are
obtained.

Excluding the contribution of the bright Be star, the broad-band
spectral energy distribution of {\lsi} peaks in the MeV-GeV band (see,
e.g., Fig.~6 in \citealt{chernyakova2006}, and Fig.~2 in
\citealt{gupta2006}, and references therein). The 0.1--300 GeV flux
observed by {\textsl Fermi} Large Area Telescope when the NS is at
apastron was recently estimated by \citet{hadasch2012} as
$5.1(3)\times10^{-10}$ erg cm$^{-2}$ s$^{-1}$, corresponding to a
luminosity of $\sim2.5\times10^{35}$ d$_{2}^2$ erg s$^{-1}$, where
d$_{2}$ is the distance to the source in units of 2 kpc. Such a value
is $\sim20\%$ lower than the flux observed when the compact object is
close to periastron. A flux of $\sim2\times10^{-10}$ was observed by
COMPTEL on-board the {\textsl Compton Gamma-Ray Observatory}, in the
1--10 MeV band \citep{tavani1996}, while the source is significantly
dimmer in the X-ray and TeV energy bands \citep[see, e.g.][and
  references therein]{hadasch2012}.

If such luminosities are ejector-only generated, the spin-down power
of the NS must lie in a range between ${\rm few}\times10^{35}$ and
${\rm few}\times10^{37}$ erg s$^{-1}$ \citep[see, e.g., the discussion
  of][ who considered an efficiency of spin-down to $\gamma$-ray
  luminosity conversion of 0.034($L_{\rm ej}/10^{36}$ erg
  s$^{-1}$)$^{-1/2}$, following \citealt{abdo2010}]{zabalza2011}; the
lower end of this interval is obtained when the emission is beamed in
a 1~sr solid angle, while the higher values correspond to more
isotropic pulsar beaming models.  It is then clear how a pulsar with
such an ejector luminosity is unlikely to lie to the right of the red
line labeled as $10^{35}$ in Fig.~\ref{fig:BPregimes}, as well as to
the left of the line labeled as $10^{37}$.  Even recalling how the
displacement of the tracks plotted in Fig.~\ref{fig:BPregimes} depends
on the exact values of unknown parameters, such as the maximum rate of
mass captured by the NS, as well as on the assumed ejector braking law
and details on the ejector/propeller transition, the relatively large
gamma-ray luminosity indicates that the NS must be young, and close to
the transition between the ejector phase and the {\it flip-flop}
state.

Thus, if the source lies in a {\it flip-flop} state, its period must
be close to the value it had at the beginning of such a phase,
i.e. that given by Eq.~(\ref{eq:Plow}). This obviously reduces the
time spent in the {\it flip-flop} state before reaching a period
corresponding to a certain ejector luminosity (for a given magnetic
field), with respect to that given by Eq.~(\ref{eq:flipfloptime}). For
the case of a maximum mass capture rate of $\dot{m}_1^{max}=1$ and a
magnetic field $b_1=1$, the time spent in such a state before spinning
down to a period corresponding to an ejector luminosity of
$5\times10^{35}$ erg s$^{-1}$, is $3.1$ kyr.  By considering larger
values of the ejector luminosity, the range of periods for which the
NS is in the {\it flip-flop} phase further reduces (see
Fig.~\ref{fig:BPregimes}); a timescale of 1.8 kyr is obtained for an
ejector luminosity of $10^{36}$ erg s$^{-1}$, while extremely small
values, $\simlt0.1$ kyr, are obtained if $L_{ej}\simgt
2.5\times10^{36}$ erg s$^{-1}$.  From Fig.~\ref{fig:BPregimes} it is
also clear that the interval of periods for which the ejector
luminosity deduced from the observed $\gamma$-ray flux is compatible
with a {\it flip-flopping} behavior increases for lower magnetic
fields and larger maximum mass accretion rates, with respect to the
fiducial values considered.

Alternatively, the luminosity observed when the source is close to the
apastron has been interpreted as powered by the propeller
emission \citep[e.g.,][]{bednarek2009}. In such a case {\lsi} would
lie at the right of the {\it flip-flop}/propeller transition (green
area of Fig.~\ref{fig:BPregimes}). However, even considering the
stronger propeller torque introduced,
Eq.~(\ref{ghosh}), the expected luminosity is:
\begin{eqnarray}
& L_{\rm prop}& =  \Omega N_{\rm prop}^G\nonumber \simeq\\ &&
\!\!\!\!\!\!\!\!\!\!\!\!\!\!\!\!
\simeq 0.9\times10^{36}\:b_1^{8/7}\:\dot{m}_1^{3/7}\:r_1^{24/7}\:m_{1}^{-2/7}\:P^{-2}\:{\rm erg}\:{\rm s}^{-1}  
\label{eq:propem}
\end{eqnarray} where $P$ is the NS spin period in seconds.
Such a luminosity is then compatible with the luminosity observed from
the system when the NS is close to periastron (i.e. for
$\dot{m}_1\simeq1$), if the period is $\simlt 1$ s, i.e. if the system
is close to the transition between the ejector and the {\it flip-flop}
phase. On the other hand, if the spin period is $\simgt 4$ s, as it is
implied by the assumption that the source is always in the propeller
state, such an effect is barely capable to power the observed emission
close to periastron ($\dot{m}_1\simeq1$), for a NS magnetic field
$\simgt 10^{14} $ G (i.e. $b_1\simgt10$). Close to apastron instead,
the rate of mass capture is much lower, $\dot{m}\sim 10^{-4}$, and the
propeller falls short by $\sim$ two order of magnitudes in accounting
for the observed $\gamma$-ray luminosity (but see below).

While the timescales derived so far do not depend much on the value of
the minimum mass capture rate, $\dot{m}_1^{min}$, this parameter plays
a major role in determining the state of the NS at orbital phases
close to apastron, and therefore the value of the period at which the
system abandons the {\it flip-flop} state (i.e. the location of the track
labeled as {\it flip-flop}/propeller in Fig.~\ref{fig:BPregimes}).  In
particular, an increase of the rate of mass captured by the NS,
possibly caused by an increase of the rate of mass lost by the Be star
and/or by a growth of the size equatorial disc, may bring the system
out of the {\it flip-flop} state.  The magnitude of the variation of the
mass capture rate needed to determine such a transition obviously
depends on the initial location of the source in a diagram like that
of Fig.~\ref{fig:BPregimes}. If the source lies close the {\it
  flip-flop}/propeller transition, a variation by a factor of few of
the mass capture rate at apastron, owing for instance to an
enhancement of the rate of mass lost by the Be star through its polar
wind, is sufficient to inhibit the ejector emission throughout the
orbit. On the other hand, if the source lies close to the ejector/{\it
  flip-flop}, as it is deduced from the observed $\gamma$-ray
luminosity, a much larger increase of the rate of mass captured at
apastron is needed to obtain such a transition. This is in principle
feasible if an increment of the rate of mass lost by the Be star is
accompanied by an increase of the radius of the equatorial disc, so to
exceed the maximum orbital separation (see the black dashed line in
Fig.~\ref{fig:mdot}). In this framework, the super-orbital variability
hinted at in TeV \citep{li2012} would be understood in terms of a
transition towards the propeller state at all orbital phases, due to
an increase of the mass loss rate of the Be star and by an increase of
its size. Such a significant increase of the mass capture rate when
the NS is close to apastron would be also able to explain the observed
$\gamma$-ray luminosity in terms of the propeller emission [see
  Eq.~(\ref{eq:propem})].

\section{Discussion}
\label{sec:discussion}

According to the conventional picture \citep[see,
  e.g.][]{lipunov1992}, a NS in a binary system evolves through
ejector, propeller and accretor states, as it spins down. While on
long timescales the transition between these states is mainly set by
the evolution of the spin period of the NS, on shorter timescales a
key role is played by the rate at which the compact object captures
the mass lost by the companion star. In particular, the large values
of the ratio between the maximum and minimum rate of mass captured by
the NS achieved if the orbit is eccentric, may induce state
transitions along an orbital cycle, a {\it flip-flop} state. The range
of mass capture rates spanned through an orbital cycle is even larger
if the non degenerate star is of Be class, losing mass also through a
dense equatorial disc which is transversed by the NS when it is close
to periastron. The existence of systems alternating states on the
timescale set by the orbital period is therefore a natural consequence
of the way a NS evolves.

While the observation of two magnetar-like bursts from a few-arcmin
region compatible with the position of {\lsi} provided a strong
indication in favor of a NS nature of the compact object in the
system, it is unlikely that an accreting NS is hosted by {\lsi}.
Accreting NS in Be-HMXB show in fact X-ray pulsations at a
period larger than few seconds. Moreover, their X-ray energy continuum
spectrum is described by a power law with an exponential cutoff at
10--30 keV, on which cyclotron and/or iron emission features are
generally superimposed \citep[see][and references
  therein]{reig2011}. No X-ray pulsation has been detected so far from
{\lsi} \citep{rea2010}, while its X-ray spectrum is featureless and
does not show a cut-off in the X-ray energy band
\citep{chernyakova2006,zhang2010}.

On the other hand, the luminosity observed from the system is of the
same order of the spin-down power liberated by the NS when its spin
period is close to the critical value marking the transition from the
ejector to the {\it flip-flop} phase.  An important clue to estimate
the likelihood of finding a NS in {\lsi} in such a phase follows from
the assessment of the time spent by the system in such a state. We
showed how such a timescale crucially depends on the details of the
assumed propeller torque. By assuming the rotating NS to release to
the incoming matter the energy needed to unbind it (see Eq.~\ref{is}),
or the energy effectively stored in the NS rotation (Eq.~\ref{ghosh}),
largely different estimates of the evolutionary timescales are
obtained.  This is essentially because the rotational energy of the NS
when spinning close to the ejector/propeller transition largely
exceeds the gravitational energy possessed by the in-falling matter,
evaluated at the large magnetospheric radius implied by a strong
magnetic field. Among the propeller torques considered here, the
weaker yields a total duration of the {\it flip-flop} phase which is
much larger ($\approx 280$ kyr) than the one implied by the stronger
torque ($\approx 25$ kyr).  Even considering the stronger propeller
torque, which is favored if the system is powered by the propeller
effect when the NS is close to periastron, the expected total duration
of the {\it flip-flop} phase [see Eq.~(\ref{eq:flipfloptime})] is
larger by a factor of $\approx 4$ with respect to the timescale spent
by the object in the pure ejector state:
\begin{equation}
\frac{t_{ff}}{t_{ej}}\approx 4 \:b_1^{-0.24}\: (\dot{m}_1^{max})^{0.27}
\label{eq:ratio}
\end{equation}
where we have made explicit only the dependence of
Eq.~(\ref{eq:emtime}) and (\ref{eq:flipfloptime}) on the NS magnetic
field, and on the maximum mass accretion rate.  It is then reasonable
to find the system in the {\it flip-flop} state. Moreover, even if it
is taken into account that the system must be relatively young to emit
a spin-down luminosity $\simgt{\rm few}\times10^{35}$ erg s$^{-1}$,
the time spent by the system in the {\it flip-flop} phase is of few kyr,
comparable to the total duration of the previous ejector phase, and
yielding an age of the system $\approx 5$--$10$ kyr, of the order of
typical spin-down ages of magnetars. On the other hand, a spin-down
luminosity of the order of $\approx10^{37}$ erg s$^{-1}$, would imply
a smaller age for the system, with the NS emitting as an ejector all
along the orbit.

\begin{deluxetable}{lccccc}
\tablecaption{Ante-diluvian systems and {\lsi}\label{tab:antediluvian}} 
\tablehead{
  \colhead{Name} & \colhead{$P_S$(s)} & \colhead{$P_{orb}$(d)}  & \colhead{$e$}  & \colhead{$B_1$(G)}\tablenotemark{(a)} & $M_2$ ($M_{\odot}$) } \startdata 
J1740-3052 & 0.57 & 231.0   & 0.57 & 3.9$\times10^{12}$ & 11.0--15.8 \\
J1638-4725 & 0.76 & 1941 & 0.95 & 1.9$\times10^{12}$ & 5.8--8.1 \\
J0045-7319& 0.93 &  51.1 &   0.81 & 2.1$\times10^{12}$ & 3.9--5.3 \\ 
B1259-63    &  0.048 & 1237 & 0.87 & 3.3$\times10^{11}$ & 3.1--4.1 \\
& & & & & \\
{\lsi}& \nodata & 26.5 & 0.63 & \nodata & 10--15 
 \enddata
 \tablenotetext{(a)}{The magnetic field of pulsars is determined as $3.2\times10^{19}\:(P\dot{P})^{1/2}$ G.}
\tablerefs{ \citet{mcconnell1991, johnston1992,kaspi1996a,stairs2001, wang2004,casares2005,lorimer2006}; we acknowledge the use of the ATNF Pulsar Catalogue, http://www.atnf.csiro.au/people/pulsar/psrcat/, \citet{manchester2005}}
\end{deluxetable}

Either it lying in the ejector or in the {\it flip-flop} state, the
presence of a young NS in {\lsi} would make the system closely related
to so-called {\it ante-diluvian systems} \citep{vandenheuvel2001},
rotation powered pulsars orbiting a high mass companion in a eccentric
orbit, considered as the progenitors of HMXB.  The properties of the
four sources of this class discovered so far are listed in
Table~\ref{tab:antediluvian}, together with those of {\lsi}. We note
that B1259-63 is also one of the brightest $\gamma$-ray binaries
known. All these NS have a superficial magnetic field in the range
$3\times10^{11}$--$4\times10^{12}$ G, as derived from the observed
spin down rate. Indeed, it was early proposed that some of these
systems could be found in a propeller state when the NS was close to
periastron. However, the rate at which the companion star would have
to lose mass in order to overcome the pulsar pressure should be much
larger than expected \citep[][and references
  therein]{campana1995,ghosh1995a,tavani1997} or observed
\citep{kaspi1996b}.  Despite a proper evaluation of the ratio defined
by Eq.~(\ref{eq:ratio}) for the known {\it ante-diluvian} systems is
far from the scope of this paper, it can be noted how the likelihood
of observing a system in the pure ejector state increases when the
maximum mass capture rate decreases, as it is the case of at least
three out of the four known {\it ante-diluvian} systems, lying in a
much larger orbit with respect to that of {\lsi}.

If confirmed by the discovery of pulsations, {\lsi} would be the first
magnetar to be discovered in a binary system. The large luminosity
variation shown by super fast X-ray transients on short timescales led
\citet{bozzo2008} to argue how a magnetar-like magnetic field could
represent an efficient gating mechanism to make these systems to
rapidly switch between propeller and accreting states. That a number
of accreting HMXB should have host in the past a NS with a field in
the magnetar-range has also been claimed on the basis of their very
large spin periods (e.g. the cases of 2S 0114+650 with a period of 2.7
hr, \citealt{li1999}, and 4U 2206+54 with a period of 5560 s,
\citealt{finger2010,ikhsanov2010,reig2012}). The spin period at which
a system eventually starts accreting mass increases in fact with the
strength of the magnetic field (see Eq.~[\ref{eq:pmax}] and the
similar expression derived by \citealt{shakura2012} for the
equilibrium period of NS in the settling regime), essentially because
the value of the magnetic field sets the strength of the propeller
torques that are responsible for the NS spin down. In this context the
case of the Be/X-ray binary in the Small Magellanic Cloud, SXP 1062,
with a period of 1062 s, and with an estimated age of 16 kyr
\citep{haberl2012} is noteworthy. \citet{popov2012} showed how such a
short age strongly points to the presence of a NS with a large initial
magnetic field in the system, $\sim 10^{14}$ G, which could have spun
the NS down very efficiently before the start of the accretion
phase. A comparison of the age proposed for SXP 1062 with the
timescales of the different propeller mechanism listed in
Table~\ref{tab}, indicates how the faster and stronger expression for
the propeller torque is probably closer to the torque effectively
experienced by a NS in that system during the propeller stage. How a
{\it strong} propeller torque, of the order of that given by
Eq.~(\ref{ghosh}), should be possibly preferred in describing the
evolution of systems like {\lsi}, is also indicated by the ratio
between the ejector and {\it flip-flop} timescales defined by
Eq.~(\ref{eq:ratio}). If the weaker propeller torques were in place, a
NS would spend a much larger time in the {\it flip-flop} state than in a
pure ejector state, and this is not indicated by the number of systems
observed in the latter state (4) with respect to the single possible
case of {\lsi}.

We finally note that, contrary to the case of SXP 1062, no association
with a supernova remnant could be been made for {\lsi}
\citep{frail1987}. This is not entirely surprising for a source with
an estimated age between 10 and 20 kyr, since such an association can
be found only for a fraction 0.64 and 0.55 of the radio pulsars with
such ages estimated from their electromagnetic spin down, respectively
(ATNF pulsar database, \citealt{manchester2005}). Furthermore, no
other Be X-ray binary has been observed embedded in a SNR, despite the
relatively young age and large number of observed systems. This is
most probably due to the large wind of the two progenitor massive
stars, which has swept away most of the material around the binary,
resulting in an under-luminous (hence undetectable) SNR after the
explosion.

\acknowledgments

This work was supported by the grants AYA2009-07391 and SGR2009-811,
as well as the Formosa program TW2010005 and iLINK program
2011-0303. We are grateful to Rosalba Perna and Jose Pons for useful
discussions.

\bibliography{biblioLSI}
\end{document}